\RequirePackage{fix-cm}
\documentclass[smallextended]{svjour3}       
\smartqed  
\usepackage{graphicx}
\usepackage{multicol}
\usepackage{xcolor}
\journalname{General Relativity and Gravitation}
\begin{document}

\title{Photon helicity and quantum anomalies in curved spacetimes}

\author{Matteo Galaverni         \and
       Gabriele Gionti, S.J. 
}


\institute{Matteo Galaverni 
           \at
           Specola Vaticana, V-00120 Vatican City, Vatican City State\\
             INAF/OAS  Bologna,  Osservatorio  di  Astrofisica  e  Scienza  dello  Spazio, via  Gobetti  101,  I-40129  Bologna,  Italy.\\
            ORCID: 0000-0002-5247-9733\\
            \email{matteo.galaverni@gmail.com}
           \and
          Gabriele Gionti, S.J. 
          \at
          Specola Vaticana, V-00120 Vatican City, Vatican City State\\
          Vatican Observatory Research Group Steward Observatory, The University Of Arizona, 
          933 North Cherry Avenue, Tucson, Arizona 85721, USA\\
          INFN, Laboratori Nazionali di Frascati, Via E. Fermi 40, 00044 Frascati, Italy\\
          ORCID: 0000-0002-0424-0648\\
          \email{ggionti@specola.va}
}

\date{Received: date / Accepted: date / Version: \today}

\maketitle

\begin{abstract}
We compare several definitions of photon helicity 
(magnetic, electric, electromagnetic)
present in literature. 
In curved spacetime quantum anomalies can spoil helicity conservation 
inducing - according to these definitions - different effects on photons:
either rotation of linear polarization or production of circular polarization. 
We derive the Noether current associated with duality transformations starting from manifestly invariant Lagrangians in Minkowski spacetime.
\end{abstract}

\section{Introduction}

In absence of charges and currents Maxwell equations are invariant under rotations of
electric and magnetic fields into each others (electromagnetic duality).
This invariance is associated with the conservation of polarization
properties of electromagnetic waves during propagation in free space.
Already in the Sixties Lipkin \cite{Lipkin1964} 
and Calkin \cite{Calkin1965} discussed the physical 
implications of this symmetry: for a free electromagnetic field the difference between the number of 
right and left circularly polarized photons is constant in time.

Quantum effects may induce violation of helicity conservation for photons propagating in curved spacetimes. After the first paper on this topic
by Dolgov, Khriplovich and Zakharov \cite{Dolgov:1987yp} several publications appeared in the Eighties 
\cite{Dolgov:1988qx,Vainshtein:1988ww,Dolgov:1987xv,Reuter:1987ju,Reuter:1991cb}.  
They discussed possible observational consequences of electromagnetism  quantization in curved spacetimes focusing  on 
\emph{the difference between the number of left-handed and right-handed photons} \cite{Dolgov:1988qx} and on \emph{the rotation of the plane of linear polarization} \cite{Reuter:1991cb}. 
More recently several papers \cite{Agullo:2014yqa,Agullo:2016lkj,Agullo:2017pyg,Agullo:2018nfv,Agullo:2018iya,delRio:2020cmv}
studied the breaking of the classical symmetry focusing the production of circular polarization in curved spacetimes.

Section II is dedicated to the comparison of the various definitions present in literature for the photon helicity: magnetic, electric and electromagnetic. 
We discuss the time dependence of the helicity densities for
photons propagating in free space and the connection with the Stokes parameters 
(for a plane monochromatic wave and for the superposition of two waves of different frequencies). 
In Section III we examine the consequences of quantum anomalies in curved spacetimes:
observable effects depend whether we focus on magnetic or electromagnetic helicity.
If magnetic helicity is not conserved the main effect is a rotation of the plane on linear polarization.
On the other side if electromagnetic helicity varies during propagation the degree of circular polarization is modified.
We discuss in detail this two different effects in the case of the Kerr metric.
In Section IV we confront and contrast various derivations of the Noether current associated with duality transformations starting from manifestly invariant Lagrangians.
We conclude in Section V.

\section{Maxwell equations and photon helicity}

\subsection{Minkowski spacetime}

In absence of sources the Maxwell equations in a non conductive medium are \cite{Jackson:1998nia}:  
\begin{eqnarray}
\label{Max:hom}
\nabla\cdot \mathbf{B}=0\,,
& &\nabla\times  \mathbf{E}+ \frac{\partial\mathbf{B}}{\partial t} =0\,,\\
\label{Max:inhom}
\nabla \cdot \mathbf{D}=0\,,
& &\nabla\times  \mathbf{H}- \frac{\partial\mathbf{D}}{\partial t} =0\,,
\end{eqnarray}
where electric field $\mathbf{E}$ and magnetic field $\mathbf{H}$ are related to electric induction $\mathbf{D}$ and magnetic induction $\mathbf{B}$ by the constitutive relations of the medium (e.g. in vacuum $\mathbf{D}=\epsilon_0\mathbf{E}$ and $\mathbf{B}=\mu_0\mathbf{H}$).\\

We have decided, following \cite{Deser:1976iy,Barnett:2012,Cameron:2014optical,Cameron:2014second,Cameron:2014Noether},
to introduce two potentials $\mathbf{A}$ and $\mathbf{C}$ relative, respectively, to magnetic induction $\mathbf{B}$  
and electric induction $\mathbf{D}$. 
\begin{multicols}{2}
\setlength{\columnseprule}{0.5pt}
\noindent
In terms of the magnetic potential $\mathbf{A}$:
  \begin{equation}
  \mathbf{B}=\nabla\times\mathbf{A}\,,
  \end{equation}

\columnbreak\noindent
In terms of the electric potential  $\mathbf{C}$:
  \begin{equation}
  \mathbf{D}=-\nabla\times\mathbf{C}\,,
  \end{equation}
\end{multicols}

\begin{multicols}{2}
\setlength{\columnseprule}{0.5pt}
\noindent
using Coulomb gauge condition $\nabla\cdot \mathbf{A}=0$, the scalar part of the potential is set equal to zero, therefore:
  \begin{equation}
  \mathbf{E}=-\frac{\partial\mathbf{A}}{\partial t}\,.
  \end{equation}

\columnbreak
\noindent
using Coulomb gauge condition $\nabla\cdot \mathbf{C}=0$, the scalar part of the potential is set equal to zero, therefore:
  \begin{equation}
  \mathbf{H}=-\frac{\partial\mathbf{C}}{\partial t}\,.
  \end{equation}
\end{multicols}

The covariant formulation is easily obtained once we have introduced the magnetic and electric potential four vectors $A^\alpha=(0,\mathbf{A})$ and $C^\alpha=(0,\mathbf{C})$. 
We define the differential two-form $F$, the electromagnetic field strength, as $F\equiv dA$ ($A$ being the four-dimensional 1-form associate to $\mathbf{A}$). 
Analogously, we introduce the differential two-form $G$ as $G\equiv dC$ 
($C$ being a 1-form associated to $\mathbf{C}$). 
This choice is done since the duality transformations has to be implemented on the true dynamical variables cfr. \cite{Deser:1976iy}. 
If one performs the duality transformations on the fields, derivatives of potentials, will incur in some technical problems (see Section IV). 
Therefore we obtain, following \cite{Cameron:2014Noether}:
\begin{eqnarray}
F_{\mu \nu} &=& \partial_\mu A_\nu - \partial_\nu A_\mu =\left( 
\begin{array}{cccc}
  0 & -E_x/c & -E_y/c & -E_z/c\\
  E_x/c & 0 &  B_z & -  B_y\\
  E_y/c & -B_z & 0 & B_x\\
  E_z/c & B_y & -B_x & 0
\end{array}
\right)\,,\\
G_{\mu \nu} &=& \partial_\mu C_\nu - \partial_\nu C_\mu =\left( 
\begin{array}{cccc}
  0 & -H_x/c & -H_y/c & -H_z/c\\
  H_x/c & 0 &  -D_z & D_y\\
  H_y/c & D_z & 0 & -D_x\\
  H_z/c & -D_y & D_x & 0
\end{array}
\right)\,,
\end{eqnarray}
where $c\equiv1/\sqrt{\mu_0\epsilon_0}$.
The Hodge dual is defined as  $^*F_{\mu\nu} \equiv \frac{1}{2} \epsilon_{\mu\nu\rho\sigma} F^{\rho\sigma}$, 
here $\epsilon_{\mu\nu\rho\sigma} \equiv \sqrt{-g}[\mu\nu\rho\sigma]$  
is the Levi-Civita symbol and $[\cdots]$ guarantees complete anti-symmetrization in the four indexes
\cite[pp. 88.97]{Misner:1974qy} \cite[p. 174]{Dubrovin:1992}:
\begin{eqnarray}
^*F_{\mu\nu} = \left( 
\begin{array}{cccc}
  0 & B_x & B_y & B_z\\
  -B_x & 0 &  E_z/c & -E_y/c\\
  -B_y & -E_z/c & 0 & E_x/c\\
  -B_z & E_y/c & -E_x/c & 0
\end{array}
\right)\,,
\\ 
^*G_{\mu \nu} =\left( 
\begin{array}{cccc}
  0 & -D_x & -D_y & -D_z\\
  D_x & 0 &  H_z/c & -H_y/c\\
  D_y & -H_z/c & 0 & H_x/c\\
  D_z & H_y/c & -H_x/c & 0
\end{array}
\right)\,.
\end{eqnarray}
If we assume propagation in in vacuum ($\mathbf{D}=\epsilon_0\mathbf{E}$ and $\mathbf{B}=\mu_0\mathbf{H}$) there is a simple relation between $^*F_{\mu\nu} $ and $G_{\mu\nu}$ ($^*G_{\mu\nu} $ and $F_{\mu\nu}$):
\begin{multicols}{2}
\setlength{\columnseprule}{0.5pt}
\noindent
\begin{equation}\label{*F}
^*F_{\mu\nu} =-\sqrt{\frac{\mu_0}{\epsilon_0}} G_{\mu\nu}\,,
\end{equation}
\columnbreak
\noindent
\begin{equation}\label{*G}
^*G_{\mu\nu}=\sqrt{\frac{\epsilon_0}{\mu_0}}F_{\mu\nu}\,,
\end{equation}
\end{multicols}

The following invariants can be explicitly evaluated:
\begin{eqnarray}
F_{\mu\nu}\,F^{\mu\nu}=-^*F_{\mu\nu}\,^*F^{\mu\nu}=2\left(\mathbf{B}^2-\frac{\mathbf{E}^2}{c^2}\right)\,,\\
F_{\mu\nu}\,^*F^{\mu\nu}=4\,\mathbf{B}\cdot\frac{\mathbf{E}}{c} \,,\\
G_{\mu\nu}\,G^{\mu\nu}=-^*G_{\mu\nu}\,^*G^{\mu\nu}=2\left(\mathbf{D}^2-\frac{\mathbf{H}^2}{c^2}\right)\,,\\
G_{\mu\nu}\,^*G^{\mu\nu}=-4\,\mathbf{D}\cdot\frac{\mathbf{H}}{c} \,.
\end{eqnarray}

\begin{multicols}{2}
[The Maxwell equations in terms of the potential are:]
\setlength{\columnseprule}{0.5pt}
\noindent
once we have introduced the potential $\mathbf{A}$ the first two Maxwell equations (Eqs.~\ref{Max:hom}) are identically verified: 
$$dF=d(dA)=0\,,$$ 
corresponding to the Bianchi identity:
$$\partial_{[\alpha}F_{\mu\nu]}=0\,,\mathrm{or: }\,\partial^\mu\,^*F_{\mu\nu}=0\,. $$
The other two equations (Eqs.~\ref{Max:inhom}) are:
$$\partial_\mu\,F^{\mu\nu}=0\,. $$
or $d\,^*F=0$ using differential forms. Corresponding to the Lagrangian density:
$$\mathcal{L}_\mathtt{F}=-\frac{1}{4\mu_0}F_{\mu\nu}F^{\mu\nu}\,.$$

\columnbreak
\noindent once we have introduced the potential $\mathbf{C}$ the last two Maxwell equations (Eqs.~\ref{Max:inhom}) are identically verified: 
$$dG=d(dC)=0\,,$$ 
corresponding to the Bianchi identity:
$$\partial_{[\alpha}G_{\mu\nu]}=0\,,\mathrm{or: }\,\partial^\mu\,^*G_{\mu\nu}=0\,. $$
The other two equations (Eqs.~\ref{Max:hom}) are:
$$\partial_\mu\,G^{\mu\nu}=0\,. $$
or $d\,^*G=0$ using differential forms. Corresponding to the Lagrangian density:
$$\mathcal{L}_\mathtt{G}=-\frac{1}{4\epsilon_0}G_{\mu\nu}G^{\mu\nu}\,.$$
\end{multicols}

In analogy with with fluid mechanics and plasma physics different gauge invariants pseudoscalar quantities can be introduced. They describe the  linking numbers of the magnetic and electric lines \cite{Trueba:1996a}, 
the knottedness of the vortex lines \cite{Barnett:2012,Crimin:2019a,Poulikakos:2019a}.
For example, in  fluid dynamics, if $\mathbf{v}$ is a vector field describing the fluid flow velocity, $\nabla\times  \mathbf{v}$ is the vorticity, and $\mathbf{v}\cdot \left(\nabla\times  \mathbf{v}\right) $ describes the knottedness of the vortex lines.
For the electromagnetic field the quantities more frequently used are:

\begin{itemize}
\item {\it Magnetic helicity} is defined using the vector potential $\mathbf{A}$:
\begin{eqnarray}
\mathcal{H}_\mathtt{mag}&\equiv&\int_{\mathbf{R}^3} h^0_\mathtt{mag} d^3\mathbf{x}\nonumber\\
&=&\frac{1}{2}\sqrt{\frac{\epsilon_0}{\mu_0}} \int_{\mathbf{R}^3} \mathbf{A}\cdot \left(\nabla\times  \mathbf{A}\right)  d^3\mathbf{x}
=\frac{1}{2}\sqrt{\frac{\epsilon_0}{\mu_0}} \int_{\mathbf{R}^3} A^i\,^*F_{0i} d^3\mathbf{x}\,.
\label{h:mag:flat}
\end{eqnarray}
\item {\it Electric helicity} is defined in terms of the vector potential $\mathbf{C}$:
\begin{eqnarray}
\mathcal{H}_\mathtt{el}&\equiv&\int_{\mathbf{R}^3} h^0_\mathtt{el} d^3\mathbf{x}\nonumber\\
&=&\frac{1}{2}\sqrt{\frac{\mu_0}{\epsilon_0}} \int_{\mathbf{R}^3} \mathbf{C}\cdot \left(\nabla\times  \mathbf{C}\right)  d^3\mathbf{x}
=\frac{1}{2}\sqrt{\frac{\mu_0}{\epsilon_0}} \int_{\mathbf{R}^3} C^i\,^*G_{0i} d^3\mathbf{x}\,.
\end{eqnarray}
\item Summing this two terms we obtain the definition of the {\it electromagnetic helicity}:
\begin{eqnarray}
\mathcal{H}_\mathtt{em} &\equiv& \int_{\mathbf{R}^3} \left(h^0_\mathtt{mag}+h^0_\mathtt{el}\right)  d^3\mathbf{x}\nonumber\\
&=&
\frac{1}{2} \int_{\mathbf{R}^3} \left[\sqrt{\frac{\epsilon_0}{\mu_0}}\mathbf{A}\cdot \left(\nabla\times  \mathbf{A}\right)
+ \sqrt{\frac{\mu_0}{\epsilon_0}}  \mathbf{C}\cdot \left(\nabla\times  \mathbf{C}\right) \right]  d^3\mathbf{x}\\
&=&
\frac{1}{2}\int_{\mathbf{R}^3} \left(\sqrt{\frac{\epsilon_0}{\mu_0}}A^i\,^*F_{0i}+C^i\, F_{0i} \right) d^3\mathbf{x}\label{h:em:flat:a}\\
&=&
\frac{1}{2}\int_{\mathbf{R}^3} \left(-A^i\,G_{0i}+ \sqrt{\frac{\mu_0}{\epsilon_0}} C^i\,^*G_{0i} \right)  d^3\mathbf{x}\label{h:em:flat:b}\\
&=&
\frac{1}{2}\int_{\mathbf{R}^3} \left(-A^i\,G_{0i}+ C^i\,F_{0i} \right)  d^3\mathbf{x}\\
&=&
\frac{1}{2}\int_{\mathbf{R}^3} \left(A_i\,G^{0i}- C_i\,F^{0i} \right)  d^3\mathbf{x}\,.
\label{h:em:flat}
\end{eqnarray}
\end{itemize}

In order to clarify the physical meaning of these quantities
we compare them to the Stokes parameters for some simple electromagnetic waves. For a {\em plane monochromatic wave} propagating in $\mathbf{z}$ direction with electric field:
\begin{equation}
\mathbf{E}=\left(E_+ e^{i\delta_+} \frac{\mathbf{x}+i\mathbf{y}}{\sqrt{2}} +
E_- e^{i\delta_-} \frac{\mathbf{x}-i\mathbf{y}}{\sqrt{2}}\right) \exp\left(i k z-i \omega t \right) \,,
\end{equation} 
here $k=\sqrt{\mu_0\epsilon_0 }\omega$. We can easily evaluate the  Stokes parameters, describing the polarization properties: 
\cite{Jackson:1998nia}[Sect 7.2] \cite{Mandel:1995}[p.348]:
\begin{eqnarray}
I&\equiv&\left\langle E_x^*(t)E_x(t)\right\rangle+\left\langle E_y^*(t)E_y(t)\right\rangle=E_+^2+E_-^2\,, \\
Q&\equiv&\left\langle E_x^*(t)E_x(t)\right\rangle-\left\langle E_y^*(t)E_y(t)\right\rangle=2 E_+ E_- \cos\left(\delta_- - \delta_+\right)\,, \\
U&\equiv&\left\langle E_x^*(t)E_y(t)\right\rangle+\left\langle E_y^*(t)E_x(t)\right\rangle=2 E_+ E_- \sin\left(\delta_- - \delta_+\right)\,,\\
V&\equiv&-i\left(\left\langle E_x^*(t)E_y(t)\right\rangle-\left\langle E_y^*(t)E_x(t)\right\rangle\right)=E_+^2-E_-^2\,,
\end{eqnarray}
the plane of linear polarization has a constant orientation angle:
\begin{equation}
\alpha=\frac{1}{2}\arctan\frac{U}{Q}=\frac{\delta_- - \delta_+}{2}\,.  
\end{equation}
Starting from the definition of the electric field we derive:
\begin{eqnarray}
\mathbf{A}&=&-\frac{i}{\omega} \left(E_+ e^{i\delta_+} \frac{\mathbf{x}+i\mathbf{y}}{\sqrt{2}} +
E_- e^{i\delta_-} \frac{\mathbf{x}-i\mathbf{y}}{\sqrt{2}}\right) \exp\left(i k z-i \omega t \right) \,,\\
\mathbf{B}&=&-\frac{i k}{\omega} \left(E_+ e^{i\delta_+} \frac{\mathbf{x}+i\mathbf{y}}{\sqrt{2}} -
E_- e^{i\delta_-} \frac{\mathbf{x}-i\mathbf{y}}{\sqrt{2}}\right) \exp\left(i k z-i \omega t \right)\,,\\
\mathbf{C}&=&-\frac{k}{\mu_0\omega^2} \left(E_+ e^{i\delta_+} \frac{\mathbf{x}+i\mathbf{y}}{\sqrt{2}} -
E_- e^{i\delta_-} \frac{\mathbf{x}-i\mathbf{y}}{\sqrt{2}}\right) \exp\left(i k z-i \omega t \right)\,,
\end{eqnarray}
and evaluate the helicity densities:
\begin{eqnarray}
h_\mathtt{mag}&=&\frac{1}{2}\sqrt{\frac{\epsilon_0}{\mu_0}}\mathrm{Re}[\mathbf{A}]\cdot\mathrm{Re}\,[\mathbf{B}]= \frac{\epsilon_0 \left(E_+^2 - E_-^2 \right)}{4 \omega}=\frac{\epsilon_0}{4\omega}V\,,\\
h_\mathtt{el}&=&-\frac{1}{2}\sqrt{\frac{\mu_0}{\epsilon_0}}\mathrm{Re}[\mathbf{C}]\cdot\mathrm{Re}\,[\mathbf{D}]=\frac{\epsilon_0 \left(E_+^2 - E_-^2 \right)}{4 \omega}=\frac{\epsilon_0}{4\omega}V\,,\\
h_\mathtt{em}&=&\frac{1}{2}\left(\sqrt{\frac{\epsilon_0}{\mu_0}}\mathrm{Re}[\mathbf{A}]\cdot\mathrm{Re}\,[\mathbf{B}]
-\sqrt{\frac{\mu_0}{\epsilon_0}}\mathrm{Re}[\mathbf{C}]\cdot\mathrm{Re}\,[\mathbf{D}]  \right)\nonumber \\
&=& \frac{\epsilon_0 \left(E_+^2 - E_-^2 \right)}{2 \omega}=\frac{\epsilon_0}{2\omega}V\,.
\end{eqnarray}
Note that, in this particular case, there is a simple relation connecting the Stokes parameter $V$
describing circular polarization and helicity densities.

For more general optical fields there is not direct proportionality between helicity density and the Stokes parameter
$V$ \cite{Cameron:2014iw}. If we consider the {\em superposition of  two monochromatic waves with opposite circular polarization and different frequencies} ($\omega_1\neq\omega_2$) propagating in $\mathbf{z}$ direction with electric field:
\begin{equation}
\mathbf{E}=E_0 \exp\left(i k_1 z-i \omega_1 t \right) \frac{\mathbf{x}+i\mathbf{y}}{\sqrt{2}}
+E_0 \exp\left(i k_2 z-i \omega_2 t \right) \frac{\mathbf{x}-i\mathbf{y}}{\sqrt{2}} \,,
\end{equation} 
here $k_{1,2}=\sqrt{\mu_0\epsilon_0}\omega_{1,2} $.
The Stokes parameters are:
\begin{eqnarray}
I&=&2E_0^2\,, \\
Q&=&2E_0^2\cos \left[2\left(\Delta\omega t -\Delta\omega z/v\right)\right]  \,, \\
U&=&-2E_0^2\sin \left[2\left(\Delta\omega t -\Delta\omega z/v\right)\right]\,,\\
V&=&0\,, 
\end{eqnarray}
with $\Delta\omega\equiv\left(\omega_2-\omega_1\right)/2$; the plane of linear polarization slowly rotates
of an angle $\alpha$ with frequency $\Delta\omega$:
\begin{equation}
\alpha=\frac{1}{2}\arctan\frac{U}{Q}=-\Delta\omega\left(t-\frac{z}{c}\right)\,.
\end{equation}
Once we have evaluated:
\begin{eqnarray}
\mathbf{A}&=&- i E_0 \left( \frac{\exp\left(i k_1 z-i \omega_1 t \right)}{\omega_1} \frac{\mathbf{x}+i\mathbf{y}}{\sqrt{2}} + \frac{\exp\left(i k_2 z-i\omega_2 t \right)}{\omega_2} \frac{\mathbf{x}-i\mathbf{y}}{\sqrt{2}}\right) \,,\\
\mathbf{B}&=&-i E_0 \left( \frac{k_1 \exp\left(i k_1 z-i \omega_1 t \right)}{\omega_1}\frac{\mathbf{x}+i\mathbf{y}}{\sqrt{2}} - \frac{k_2 \exp\left(i k_2 z-i \omega_2 t \right)}{\omega_2}\frac{\mathbf{x}-i\mathbf{y}}{\sqrt{2}} \right)\,,\\
\mathbf{C}&=&-\frac{E_0}{\mu_0} \left( \frac{k_1 \exp\left(i k_1 z-i \omega_1 t \right)}{\omega_1^2}\frac{\mathbf{x}+i\mathbf{y}}{\sqrt{2}} - \frac{k_2\exp\left(i k_2 z-i \omega_2 t \right)}{\omega_2^2} \frac{\mathbf{x}-i\mathbf{y}}{\sqrt{2}}\right)\,,
\end{eqnarray}
we obtain these expressions for helicities\cite{Crimin:2019a,Mackinnon:2019}:
\begin{eqnarray}
h_\mathtt{mag}&=&\frac{\epsilon_0 E_0^2}{4}\left(\frac{1}{\omega_1}-\frac{1}{\omega_2}\right)
 \left[1+\cos\left((k_1+k_2)z-(\omega_1+\omega_2)t\right)\right]\,,\\
h_\mathtt{el}&=&\frac{\epsilon_0 E_0^2}{4}\left(\frac{1}{\omega_1}-\frac{1}{\omega_2}\right)
 \left[1+\cos\left((k_1+k_2)z-(\omega_1+\omega_2)t\right)\right]\,,\\
 h_\mathtt{em}&=& \frac{\epsilon_0 E_0^2}{2 }\left(\frac{1}{\omega_1}-\frac{1}{\omega_2}\right)
 \left[1+\cos\left((k_1+k_2)z-(\omega_1+\omega_2)t\right)\right]\\
 &=&\epsilon_0 E_0^2\left(\frac{1}{\omega_1}-\frac{1}{\omega_2}\right)
  \cos^2 \left( \frac{k_1+k_2}{2}z-\frac{\omega_1+\omega_2}{2}t\right)\,,  
\end{eqnarray}
and considering the average over time of electromagnetic helicity:
\begin{equation}
\left\langle h_\mathtt{em}\right\rangle= \frac{\epsilon_0 E_0^2}{2 }\left(\frac{1}{\omega_1}-\frac{1}{\omega_2}\right)\,.
\end{equation}
In this case, superposition of two monochromatic waves, electromagnetic helicity density is not related to the Stokes parameter $V$.

Time evolution of the helicity densities is easily obtained from the Maxwell equations \cite{Elbistan:2016khp,Elbistan:2018fkr}:
\begin{eqnarray}
\frac{\partial h^0_\mathtt{mag}}{\partial t}+\nabla\cdot\left(\sqrt{\frac{\epsilon_0}{\mu_0}}\mathbf{E}\times\mathbf{A}\right)&=& -2 \sqrt{\frac{\epsilon_0}{\mu_0}} \mathbf{E}\cdot\mathbf{B}\,,\\
\frac{\partial h^0_\mathtt{el}}{\partial t}+\nabla\cdot\left(\sqrt{\frac{\mu_0}{\epsilon_0}}\mathbf{H}\times\mathbf{C}\right)= 2 \sqrt{\frac{\mu_0}{\epsilon_0}} \mathbf{H}\cdot\mathbf{D}&=& 2 \sqrt{\frac{\epsilon_0}{\mu_0}} \mathbf{E}\cdot\mathbf{B}\,,\\
\frac{\partial h^0_\mathtt{em}}{\partial t}+\nabla\cdot\left(\sqrt{\frac{\epsilon_0}{\mu_0}}\mathbf{E}\times\mathbf{A}+\sqrt{\frac{\mu_0}{\epsilon_0}}\mathbf{H}\times\mathbf{C}\right)&=& 0\,.
\end{eqnarray}
Integrating over tridimensional space:
\begin{eqnarray}
\label{h:mag:flat:evol}
\frac{d \mathcal{H}_\mathtt{mag}}{d t}&=& -2 \sqrt{\frac{\epsilon_0}{\mu_0}} \int_{\mathbf{R}^3} \mathbf{E}\cdot\mathbf{B}\,  d^3\mathbf{x}\,,\\
\label{h:el:flat:evol}
\frac{d \mathcal{H}_\mathtt{el}}{d t}&=& 2 \sqrt{\frac{\mu_0}{\epsilon_0}} \int_{\mathbf{R}^3} \mathbf{H}\cdot\mathbf{D}\,  d^3\mathbf{x}= 2 \sqrt{\frac{\epsilon_0}{\mu_0}} \int_{\mathbf{R}^3} \mathbf{E}\cdot\mathbf{B}\,  d^3\mathbf{x}\,,\\
\frac{d \mathcal{H}_\mathtt{em}}{d t}&=&0\,,
\end{eqnarray}
we note that $\mathcal{H}_\mathtt{em}$ is constant in time, while $\mathcal{H}_\mathtt{mag}$ and $\mathcal{H}_\mathtt{el}$ are not conserved, in general.

Helicities are explicitly defined in terms of the vector potentials, however it was shown in \cite{Barnett:2012}
that only gauge-invariant transverse pieces ($\mathbf{A}^\perp$ and $\mathbf{C}^\perp$) contribute when integration over all space is performed, therefore the integrated quantities are gauge invariant.
Conservation of electromagnetic helicity has a clear physical interpretation even in quantum theory:
the number of right circularly polarized photons minus 
the number of left circularly polarized photons is conserved \cite{Calkin1965,Trueba:1996a}.

\subsection{Curved spacetime}
Electrodynamics in curved spacetime described by the metric $g_{\mu\nu}$ can be obtained replacing ordinary derivatives $\partial_\mu$ by covariant derivatives $\nabla_\mu$.
The two tensors are now defined as:
\begin{multicols}{2}
\setlength{\columnseprule}{0.5pt}
\noindent
$$ F_{\mu \nu} = \nabla_\mu A_\nu - \nabla_\nu A_\mu\,.$$
The Lagrangian density:  
\begin{equation}
\label{L0:F}
\mathcal{L}_\mathtt{F}=-\frac{1}{4\mu_0}F_{\mu\nu}F^{\mu\nu}\,,
\end{equation}
induces the following equations:
$$\nabla_\mu\,F^{\mu\nu}=0\,.$$

\columnbreak\noindent
$$G_{\mu \nu} = \nabla_\mu C_\nu - \nabla_\nu C_\mu\,.$$
The Lagrangian density:  
\begin{equation}
\label{L0:G}
\mathcal{L}_\mathtt{G}=-\frac{1}{4\epsilon_0}G_{\mu\nu}G^{\mu\nu}\,,
\end{equation}
induces the following equations:
$$\nabla_\mu\,G^{\mu\nu}=0\,.$$
\end{multicols}
Helicity densities definitions are unchanged, but the definition of the Hodge dual contains now the determinant of the metric $g$
and in order to obtain total helicity we have to integrate on a curved volume element:
\begin{equation}
\mathcal{H}\equiv\int_{\Sigma^3} h^0d^3\mathbf{r} \,.
\end{equation}
where $d^3\mathbf{r}$ is the volume element in curved tridimensional space.

\section{Quantum anomalies: helicity non-conservation in curved spacetimes}

In the Eighties - after the first study of Dolgov, Khriplovich,  and Zakharov \cite{Dolgov:1987yp} - 
several papers \cite{Dolgov:1988qx,Vainshtein:1988ww,Dolgov:1987xv,Reuter:1987ju,Reuter:1991cb}
tried to estimate the effects of electromagnetic field quantization 
in curved spacetimes. 
They focused in particular on magnetic helicity $h^\mu_\mathtt{mag}$,
see Eq.~(\ref{h:mag:flat}), deriving a relation between vacuum expectation value $\left\langle \nabla_\mu h^\mu_\mathtt{mag} \right\rangle$
and $R_{\alpha\beta\mu\nu}\,^\star R^{\alpha\beta\mu\nu}$
the Chern-Pontryagin invariant (or Hirzebruch signature); 
where $R_{\alpha\beta\mu\nu}$ is the Riemann tensor and $^*R^{\alpha\beta\mu\nu}$ its dual.

More recently Agullo {\it et al.} \cite{Agullo:2014yqa,Agullo:2016lkj,Agullo:2017pyg,Agullo:2018nfv,Agullo:2018iya,delRio:2020cmv}
focused on electromagnetic helicity $h^\mu_\mathtt{em}$,
see Eq.~(\ref{h:em:flat}).
As we remembered in the previous Section, $h^\mu_\mathtt{em}$ 
is a classical conserved current ($\nabla_\mu h^\mu_\mathtt{em}=0$), unlike $h^\mu_\mathtt{mag}$ and $h^\mu_\mathtt{em}$
which are not conserved.
They showed that the vacuum expectation value for 
$\nabla_\mu h^\mu_\mathtt{em}$ could be different from zero if the 
theory is quantized in curved metric \cite{Agullo:2018nfv}: 
\begin{equation}
\label{anomaly:density}
\left\langle \nabla_\mu h^\mu_\mathtt{em} \right\rangle = -\frac{\hbar}{96\pi^2}R_{\alpha\beta\mu\nu}\,^\star R^{\alpha\beta\mu\nu}\,.
\end{equation}
Therefore total electromagnetic helicity $\mathcal{H}_\mathtt{em}$ is not constant anymore for photons propagating in a spacetime 
with  a nonzero Chern-Pontryagin invariant:
\begin{equation}
\frac{d  \left\langle \mathcal{H}_\mathtt{em} \right\rangle}{d t} = -\frac{\hbar}{96\pi^2} \int_{\Sigma^3} R_{\alpha\beta\mu\nu}\,^\star R^{\alpha\beta\mu\nu} d^3\mathbf{r}\,.
\end{equation}
Helicity variation over a finite interval of time is proportional to:
\begin{equation}
\label{anomaly:int}
\Delta\left\langle \mathcal{H}_\mathtt{em} \right\rangle \propto
\int_{t_1}^{t_2}\int_{\Sigma^3} R_{\alpha\beta\mu\nu}\,^\star R^{\alpha\beta\mu\nu} \sqrt{-g}d^4x\,.
\end{equation}
The term ``photon chiral anomaly'' is used when 
the classical symmetry is not
conserved due to the quantization of the electromagnetic field in curved spacetime. 

In the following part of this Section we show that observable effects are different whether we focus on magnetic or electromagnetic helicity non-conservation. 

\subsection{Non-conservation of magnetic helicity}
Several papers \cite{Dolgov:1987xv,Reuter:1991cb,Campbell:1990ai,Duncan:1992vz} estimated  the  observable  effects of non-conservation of  magnetic  helicity. 
We have already shown that  is not conserved even at classical level, see Eq.~(\ref{h:mag:flat:evol}): 
\begin{equation}
\frac{d \mathcal{H}_\mathtt{mag}}{d t}\propto- \int_{\mathbf{R}^3} \mathbf{E}\cdot\mathbf{B}\,  d^3\mathbf{r}
\propto \int_{\mathbf{R}^3} F_{\mu\nu}\,^*F^{\mu\nu}\,  d^3\mathbf{r}\,.
\end{equation}
Therefore the quantization in curved spacetime induces a term 
$\left\langle F_{\mu\nu}\,^*F^{\mu\nu} \right\rangle\propto - R_{\alpha\beta\mu\nu}\,^\star R^{\alpha\beta\mu\nu}$.
When light propagates in a region with $R_{\alpha\beta\mu\nu}\,^\star R^{\alpha\beta\mu\nu}\neq 0$ 
polarization is modified since $F_{\mu\nu}\,^*F^{\mu\nu}$ acquires a nonzero vacuum expectation value. 
The effects can be estimated, following \cite{Dolgov:1987xv,Reuter:1987ju,Reuter:1991cb}, introducing a  ``conventional pseudoscalar field'' 
$\phi(x)$ coupled to photons via the term:
\begin{equation}
\mathcal{L}_\phi=\frac{1}{2} g_\phi \phi F_{\mu\nu}\,^*F^{\mu\nu}\,.
\end{equation}
In this case the main effect on the propagation of photons is the rotation of the plane of linear polarization (birefringence), 
where the angle is proportional to the variation of the pseudoscalar field $\phi(x)$ along the line of sight
\cite{Carroll:1989vb,Carroll:1991zs,Harari:1992ea}.
Also the degree of circular polarization may vary,
but the effect is subdominant
\cite{Lee:1999ae,Finelli:2008jv,Alexander:2019sqb}.

The evolution of $\phi(x)$ is obtained from the the equation of motion:
\begin{equation}
\label{phi:evol}
\nabla_\mu\nabla^\mu \phi(x)=\frac{g_\phi}{4}F_{\mu\nu}\,^*F^{\mu\nu}\propto - R_{\alpha\beta\mu\nu}\,^\star R^{\alpha\beta\mu\nu}\,.
\end{equation}
once we have specified the metric and evaluated 
$R_{\alpha\beta\mu\nu}\,^\star R^{\alpha\beta\mu\nu}$.

If we consider, for example, the Kerr Metric \cite{Ciufonini}:
\begin{eqnarray}
ds^2=&-&\left(1-\frac{2 G m r}{r^2+ a^2 \cos^2 \theta} \right) dt^2+\frac{4 G m r a \sin^2\theta}{r^2+ a^2 \cos^2 \theta} dt d\phi \nonumber\\
&+&\frac{r^2+ a^2 \cos^2 \theta}{r^2-2 G m r + a^2} dr^2  
+\left( r^2+ a^2 \cos^2 \theta \right) d\theta^2 \nonumber\\
&+& \left(r^2+a^2+\frac{2 G m r a^2 \sin^2\theta}{r^2+ a^2 \cos^2 \theta}\right) \sin^2\theta d\phi^2\,, 
\end{eqnarray}
where $a\equiv \frac{J}{m}$ is angular momentum per unit mass, we can evaluate:
\begin{eqnarray}
\mathbf{^* R}\cdot\mathbf{R}&=&-1536 a G^2 m^2 \cos \theta \left[\frac{r^5}{\left( r^2+ a^2 \cos^2 \theta \right)^{6}}-\frac{r^3}{\left( r^2+ a^2 \cos^2 \theta \right)^{5}}\right.\nonumber\\
& &\left. +\frac{3 r}{16 \left( r^2+ a^2 \cos^2 \theta \right)^4}\right]\nonumber\\
&\simeq& - 288 \frac{a G^2 m^2}{r^7} \cos \theta+\cdots\,,
\end{eqnarray}
in agreement with \cite{Ciufonini}[p. 356], and we find the following solution of Eq.~(\ref{phi:evol}):
\begin{eqnarray}
\phi(r,\theta)&=&\frac{5}{384}\frac{g_\phi \hbar}{4\pi^2}\frac{a \cos\theta}{(G m)^3}\left[\frac{4 (G m)^2}{r^2}+\frac{8 (G m)^3}{r^3}+\frac{72 (G m)^4}{5r^4}\right]\nonumber\\
&=&\frac{5}{384 \pi^2}\frac{g_\phi \hbar a}{G m}\frac{\cos\theta}{r^2} +\mathcal{O}\left(\frac{1}{r^3}\right)\,, 
\end{eqnarray}
see also Eq.~(12) of \cite{Reuter:1991cb}.  

The amount of rotation of linear polarization depends from the physical properties of the Kerr black hole
and is in general extremely small. Moreover this effect must be compared 
with the classical General Relativity effect due to photon propagation in a curved spacetime: gravitational Faraday rotation (or Skrotskii effect), 
see \cite{Piran:1985,Sereno:2004jx}. An estimate of this classical effect is given in Eq. (31) of \cite{Sereno:2004cn}:
\begin{equation}
  \Omega_{SK}^{Kerr}=-\frac{\pi}{4}\frac{G^2 m^2}{c^5}\frac{a \cos \theta}{r_{min}^3}\,.
\end{equation}

\subsection{Non-conservation of electromagnetic helicity}

Quantization of electromagnetic field in curved spacetime can spoil 
electromagnetic helicity conservation (photon chiral anomaly).
Here the observable effects of the anomaly - 
see \cite{Agullo:2016lkj,Agullo:2017pyg,Agullo:2018nfv,Agullo:2018iya} 
and in particular \cite{delRio:2020cmv} -
are directly connected to the variation of electromagnetic helicity, see Eq.~(\ref{anomaly:int}):
\begin{equation}
\left\langle \mathcal{H}_\mathtt{em}\left(t_1\right) \right\rangle - \left\langle \mathcal{H}_\mathtt{em}\left(t_2\right) \right\rangle
\propto\int_{t_1}^{t_2}\int_{\Sigma^3} R_{\alpha\beta\mu\nu}\,^\star R^{\alpha\beta\mu\nu} \sqrt{-g}d^4x\,.
\end{equation}
If the integral in the right term is different from zero, then $\mathcal{H}_\mathtt{em}$ is not conserved. 
The difference between the numbers of right circularly polarized photons
and left circularly polarized photons changes: the degree of circular polarization is not conserved.
The observable effect associated with this formulation of the quantum anomaly is not related with a change in linear polarization angle, 
but with a variation of the degree of circular polarization. 

For the particular case of the Kerr metric, discussed in the previous subsection, we have:
\begin{equation}
\int R_{\alpha\beta\mu\nu}\,^\star R^{\alpha\beta\mu\nu} \sqrt{-g}d^4x \propto 
\int_0^\pi \cos\theta \sin\theta \left[r^2+a^2\frac{1+\cos(2\theta)}{2}\right]d\theta=0\,.
\end{equation}
Therefore, since the integral over all space is zero - due to symmetry reasons - 
in this case there are no observable effects related to the quantum anomaly.

In order to have a non zero effect a metric with a nonzero Chern-Pontryagin integrated term should be considered.
In del Rio {\it et al.} \cite{delRio:2020cmv} some estimates are numerically derived for non-stationary spacetimes.

\section{Dual symmetry and photon helicity density}

In order to clarify the helicity definition that should be used 
to study quantum anomalies we compare various derivations
of the Noether current associated to invariance under duality transformations.
Several authors pointed out that transformations has to be implemented 
on the true dynamical variables cfr. \cite{Deser:1976iy,Deser:1996xp}. 
Moreover there is a further argument against duality transformations 
implemented at the level of the fields. 
In fact vector fields and pseudovectors will me mixed, 
in this way, generating inconsistencies as well \cite{Aschieri:2008ns}.
Therefore we define the transformations at the level of the potentials \cite{Cameron:2014Noether}: 
\begin{eqnarray}
\label{dual:pot}
	\mathbf{A}\rightarrow\mathbf{A}\cos\theta+\sqrt{\frac{\mu}{\epsilon}}\mathbf{C}\sin\theta\,,\;	\mathbf{C}\rightarrow\mathbf{C}\cos\theta-\sqrt{\frac{\epsilon}{\mu}}\mathbf{A}\sin\theta\,,
\end{eqnarray}
and for the tensors $F_{\mu\nu}$ and $G_{\mu\nu}$:
\begin{eqnarray}
\label{dual:FG}
	F_{\mu\nu}\rightarrow F_{\mu\nu}\cos\theta+\sqrt{\frac{\mu}{\epsilon}}G_{\mu\nu}\sin\theta\,,\;	
	G_{\mu\nu}\rightarrow G_{\mu\nu}\cos\theta-\sqrt{\frac{\epsilon}{\mu}}F_{\mu\nu}\sin\theta\,,
\end{eqnarray}
Taking the time derivative and the curl of Eqs.~(\ref{dual:pot}) we obtain the well known
 {\it duality transformations} \cite[p. 274]{Jackson:1998nia}:
\begin{eqnarray}
	\mathbf{E}\rightarrow\mathbf{E}\cos\theta+\sqrt{\frac{\mu}{\epsilon}}\mathbf{H}\sin\theta\,,\;	\mathbf{H}\rightarrow\mathbf{H}\cos\theta-\sqrt{\frac{\epsilon}{\mu}}\mathbf{E}\sin\theta\,,\\
	\mathbf{D}\rightarrow\mathbf{D}\cos\theta+\sqrt{\frac{\epsilon}{\mu}}\mathbf{B}\sin\theta\,,\;	\mathbf{B}\rightarrow\mathbf{B}\cos\theta-\sqrt{\frac{\mu}{\epsilon}}\mathbf{D}\sin\theta\,.
\end{eqnarray}
Maxwell equations, Eqs.~(\ref{Max:hom}-\ref{Max:inhom}), are manifestly invariant under such transformations for a real angle $\theta$. In vacuum  ($\mathbf{D}=\epsilon_0\mathbf{E}$ and $\mathbf{B}=\mu_0\mathbf{H}$):
\begin{eqnarray}
	\mathbf{E}\rightarrow\mathbf{E}\cos\theta+\frac{\mathbf{B}}{\sqrt{\mu_0\epsilon_0}}\sin\theta\,,\;	\mathbf{B}\rightarrow\mathbf{B}\cos\theta-\sqrt{\mu_0\epsilon_0}\,\mathbf{E}\sin\theta\\
	\label{duality:F}
	F_{\mu\nu}\rightarrow F_{\mu\nu}\cos\theta-\,^*F_{\mu\nu}\sin\theta\,,\;	
	G_{\mu\nu}\rightarrow G_{\mu\nu}\cos\theta+\,^*G_{\mu\nu}\sin\theta\,.
\end{eqnarray}
Note that for $\theta=\pi/2$ they simply exchange electric and magnetic fields:
\begin{equation}
\mathbf{E}\rightarrow\frac{\mathbf{B}}{\sqrt{\mu_0\epsilon_0}}\,,\;	\mathbf{B}\rightarrow-\sqrt{\mu_0\epsilon_0}\,\mathbf{E}\,,
\end{equation}
or a field with its dual:
\begin{equation}
F_{\mu\nu}\rightarrow -\,^*F_{\mu\nu}\,,\;	
	G_{\mu\nu}\rightarrow \,^*G_{\mu\nu}\,.
\end{equation}
It was well known, already ad the end of Nineteenth century 
(Heaviside-Larmor symmetry),
that in free space Maxwell equations  are invariant under an
exchange of $\mathbf{E}$ and $\mathbf{B}$. Later the properties under duality transformations were studied also at the level of the Lagrangian densities, Eqs.~(\ref{L0:F}-\ref{L0:G}) \cite{Calkin1965,Deser:1976iy,Deser:1982}:
\begin{eqnarray}
	\mathcal{L}_\mathtt{F}&\rightarrow& \mathcal{L}_\mathtt{F} \cos^2\theta + \mathcal{L}_\mathtt{G}\sin^2\theta-\frac{1}{2\sqrt{\mu_0\epsilon_0}}F_{\mu\nu}G^{\mu\nu}\sin\theta\cos\theta\nonumber
	\\&=& \mathcal{L}_\mathtt{F} \cos\left(2\theta\right) + \frac{1}{4 \mu_0} F_{\mu\nu}\,^*F^{\mu\nu}\sin\left(2\theta\right)\nonumber\\
	&=&\mathcal{L}_\mathtt{F} \cos\left(2\theta\right) + \frac{1}{2 \mu_0} \nabla_\mu\left(A_\nu \,^*F^{\mu\nu} \right)\sin\left(2\theta\right) \,,\\
			\mathcal{L}_\mathtt{G}&\rightarrow& \mathcal{L}_\mathtt{G} \cos^2\theta + \mathcal{L}_\mathtt{F}\sin^2\theta+\frac{1}{2\sqrt{\mu_0\epsilon_0}}F_{\mu\nu}G^{\mu\nu}\sin\theta\cos\theta\nonumber\\
			&=& \mathcal{L}_\mathtt{G} \cos\left(2\theta\right) + \frac{1}{4 \epsilon_0} G_{\mu\nu}\,^*G^{\mu\nu}\sin\left(2\theta\right)\nonumber\\
			&=&\mathcal{L}_\mathtt{G} \cos\left(2\theta\right) + \frac{1}{2 \epsilon_0} \nabla_\mu\left(C_\nu \,^*G^{\mu\nu} \right)\sin\left(2\theta\right) \,.
\end{eqnarray}

where we have used the relations $^*F_{\mu\nu}=-\sqrt{\mu_0/\epsilon_0}\,  G_{\mu\nu}$, 
$^*G_{\mu\nu}=\sqrt{\epsilon_0/\mu_0}\,F_{\mu\nu}$, Eqs.~(\ref{*F}-\ref{*G}), 
and the Bianchi identities.
Even if the Maxwell equations are invariant under duality transformations, $\mathcal{L}_\mathtt{F} $ and $\mathcal{L}_\mathtt{G} $ 
are not manifestly invariant under transformations, see Eq.~(\ref{duality:F}) \cite{Deser:1976iy,Bliokh:2012zr}. 
If we consider an infinitesimal transformation:
\begin{eqnarray}
	\mathcal{L}_\mathtt{F}&\rightarrow& \mathcal{L}_\mathtt{F} + \frac{1}{\mu_0} \nabla_\mu\left(A_\nu \,^*F^{\mu\nu} \right)\theta \,,\\
			\mathcal{L}_\mathtt{G}&\rightarrow& \mathcal{L}_\mathtt{G} + \frac{1}{\epsilon_0} \nabla_\mu\left(C_\nu \,^*G^{\mu\nu} \right) \theta \,,
\end{eqnarray}
the Lagrangians change by a total derivative term.
Sometimes it is speculated that dual invariance of the Maxwell theory should implies also the current conservation \cite[p. 24]{Shnir} \cite[p. 24]{S:2020zsp}:
\begin{eqnarray}
\nabla_\mu\left(A_\nu \,^*F^{\mu\nu} \right)	&\equiv& 0\;\Longrightarrow\;
\nabla_0 h^0_\mathtt{mag}+\nabla_i h^i_\mathtt{mag}=0\,,\\
\nabla_\mu\left(C_\nu \,^*G^{\mu\nu} \right)  &\equiv& 0\;\Longrightarrow\;
\nabla_0 h^0_\mathtt{el}+\nabla_i h^i_\mathtt{el}=0\,,
\end{eqnarray}
and therefore $\mathcal{H}_\mathtt{mag}$ and $\mathcal{H}_\mathtt{el}$ should be constant in time. 
But we have already seen in Section 2 that $\mathcal{H}_\mathtt{mag}$ and $\mathcal{H}_\mathtt{el}$ are not conserved in general,
Eqs.~(\ref{h:mag:flat:evol}-\ref{h:el:flat:evol}). 
The dual-asymmetric definitions for helicity 
are not satisfactory in the general case \cite[p. 6]{Bliokh:2012zr}.
Even if we remember that the Noether's theorem can be applied also if the action is changed by a surface term \cite[p. 18]{Peskin:1995ev} we obtain these expression for the conserved currents:
\begin{eqnarray}
h^\mu_\mathtt{F}&=&\frac{\partial\mathcal{L}_\mathtt{F}}{\partial\left(\nabla_\mu A_\nu\right)}\delta A_\nu-\frac{1}{\mu_0}A_\nu \,^*F^{\mu\nu}\nonumber\\
&=&-\frac{F^{\mu\nu}}{\mu_0}\sqrt{\frac{\mu_0}{\epsilon_0}}C_\nu-\frac{1}{\mu_0}A_\nu \,^*F^{\mu\nu}
=\frac{-1}{\sqrt{\mu_0\epsilon_0}}\left(F^{\mu\nu}C_\nu+\sqrt{\frac{\epsilon_0}{\mu_0}}A_\nu \,^*F^{\mu\nu}\right)
\\
h^\mu_\mathtt{G}&=&\frac{\partial\mathcal{L}_\mathtt{G}}{\partial\left(\nabla_\mu C_\nu\right)}\delta C_\nu-\frac{1}{\epsilon_0} C_\nu \,^*G^{\mu\nu}\nonumber\\
&=&\frac{G^{\mu\nu}}{\epsilon_0}\sqrt{\frac{\epsilon_0}{\mu_0}}A_\nu-\frac{1}{\epsilon_0}C_\nu \,^*G^{\mu\nu}
=\frac{1}{\sqrt{\mu_0\epsilon_0}}\left(G^{\mu\nu} A_\nu-\sqrt{\frac{\mu_0}{\epsilon_0}}C_\nu \,^*G^{\mu\nu}\right)\,.
\end{eqnarray}
If we restrict ourselves to Minkowski spacetime the time components are:
\begin{eqnarray}
 h^0_\mathtt{F}&=&\frac{1}{\sqrt{\mu_0\epsilon_0}}\left(F_{0i}C^i+\sqrt{\frac{\epsilon_0}{\mu_0}}A^i \,^*F_{0i}\right)=2 h^0_\mathtt{em}\,,\\
 h^0_\mathtt{G}&=&\frac{1}{\sqrt{\mu_0\epsilon_0}}\left(-G_{0i} A^i+\sqrt{\frac{\mu_0}{\epsilon_0}}C^i \,^*G_{0i}\right)=2 h^0_\mathtt{em}\,.
\end{eqnarray}
We can compare these expressions with the ones obtained in Section II for the electromagnetic helicity, see in particular Eq.~(\ref{h:em:flat:a})
and Eq.~(\ref{h:em:flat:b}). 
The factor $c=1/\sqrt{\mu_0\epsilon_0}$ is easily explained since $\frac{\partial h^0}{\partial x^0}=\frac{1}{c}\frac{\partial h^0}{\partial t}$, but here a factor $1/2$ is missing.

Already in the Seventies Deser and Teitelboim  \cite{Deser:1976iy} pointed out that a transformation is meaningful only if it can be implemented at the level of the physical variables.
More recently Refs. \cite{Agullo:2016lkj,Agullo:2018nfv} showed that the transformation exchanging
 $F$ and $\,^*F$, see Eq.~(\ref{duality:F}) 
is meaningful only when the equations of motions are both satisfied (on-shell), 
since it exchanges $\nabla_\mu \,F^{\mu\nu}=0$ with $\nabla_\mu \,^*F^{\mu\nu}=0$.
In the Lagrangian second order formalism it is not possible to use on-shell 
equations to prove that the theory is invariant under duality transformations \cite{Bernabeu:2019dym}.

Following \cite{Cameron:2014Noether,Bliokh:2012zr} we introduce the Lagrangian:
\begin{equation}
\mathcal{L}_\mathtt{dual}=\frac{\mathcal{L}_\mathtt{F}+\mathcal{L}_\mathtt{G}}{2}=-\frac{1}{8\mu_0}F_{\mu\nu}F^{\mu\nu}-\frac{1}{8\epsilon_0}G_{\mu\nu}G^{\mu\nu}\rightarrow\mathcal{L}_\mathtt{dual}\,.
\end{equation}
This Lagrangian is manifestly invariant under duality transformations for any angle $\theta$, 
and the Noether's theorem gives the expected expression for electromagnetic helicity density Eq.~(\ref{h:em:flat}).
However if we rewrite $G_{\mu\nu}G^{\mu\nu}$ using Eq.~(\ref{*F}) 
we obtain $\mathcal{L}_\mathtt{dual}=0$ \cite{Cameron:2014Noether}. 
Therefore we tried not to impose directly
 Eq.~(\ref{*F}), but
 to use the method of  method of Lagrange multipliers 
and add to the Lagrangian a new term invariant under duality transformations:
\begin{equation}
\label{L:multipl}
\mathcal{L}_\mathtt{mult}\equiv \mathcal{L}_\mathtt{dual}
+\lambda\left(\frac{\mu_0}{\epsilon_0}G_{\mu\nu}\,^*G^{\mu\nu}+F_{\mu\nu}\,^*F^{\mu\nu}\right)
\rightarrow\mathcal{L}_\mathtt{mult}\,.
\end{equation} 
The condition $\partial \mathcal{L}_\mathtt{mult}/\partial\lambda=0$ corresponds to the equation:\begin{equation}
\frac{\mu_0}{\epsilon_0}G_{\mu\nu}\,^*G^{\mu\nu}=-F_{\mu\nu}\,^*F^{\mu\nu}\,,
\label{scalar}
\end{equation}
which is verified if:
\begin{equation}
\sqrt{\frac{\mu_0}{\epsilon_0}}G_{\mu\nu}=\mp^*F_{\mu\nu}\,,\mathrm{and\,its\,dual: }
\sqrt{\frac{\mu_0}{\epsilon_0}}\,^*G_{\mu\nu}=\pm F_{\mu\nu}\,.
\label{dual}
\end{equation}
If we consider the covariant derivatives:
\begin{eqnarray}
\sqrt{\frac{\mu_0}{\epsilon_0}}\nabla_\mu G^{\mu\nu}=\mp\nabla_\mu\,^*F^{\mu\nu}\Longrightarrow\;\nabla_\mu G^{\mu\nu}=0\,,\\
\sqrt{\frac{\mu_0}{\epsilon_0}}\nabla_\mu\,^*G^{\mu\nu}=\pm\nabla_\mu\,F^{\mu\nu}\Longrightarrow\;\nabla_\mu F^{\mu\nu}=0\,,
\end{eqnarray}
where we have used the Bianchi identities. Imposing the condition $\partial \mathcal{L}_\mathtt{mult}/\partial\lambda=0$ corresponds to
verify equations of motion (on-shell solutions).
If we now apply the Noether's theorem to this Lagrangian ~(\ref{L:multipl}) we obtain the correct expression for electromagnetic helicity density $h^0_\mathtt{em}$,
Eq.~(\ref{h:em:flat}). 
The problem here is that Eq.~(\ref{scalar}) has~(\ref{dual}) as one solution, which is not unique. Therefore the equations of motion derived from~(\ref{L:multipl}) 
contain Maxwell electromagnetism as solution, but, in general, 
their solution space can be bigger than electromagnetism.\footnote{The authors are deeply grateful to Dmitri Sorokin for a very helpful discussion on this topic.}

In Lagrangian formalism, the more formal approach to derive electromagnetic helicity is to consider the Lagrangian introduced by Pasti, Sorokin, and Tonin \cite{Pasti:1995tn,Pasti:1996vs,Maznytsia:1998xw}:
\begin{equation}
\label{L:PST}
\mathcal{L}_\mathtt{PST}\equiv \mathcal{L}_\mathtt{dual}
-\frac{1}{4}\frac{1}{\partial_\tau a \partial^\tau a}\left(\partial_\mu a\right) 
\mathcal{F}^\mu \, _\rho 
\left(\partial^\rho a\right)\rightarrow\mathcal{L}_\mathtt{PST}\,,    
\end{equation}
where $a(x)$ is an auxiliary scalar field and $\mathcal{F}^\mu \, _\rho$ is:
\begin{eqnarray}
\mathcal{F}^\mu \, _\rho &\equiv&  G^{\mu\nu}G_{\nu\rho} +   G^{\mu\nu}\,^*F_{\nu\rho}+ ^*F^{\mu\nu}G_{\nu\rho} + ^*F^{\mu\nu}\,^*F_{\nu\rho}\nonumber\\
& & +F^{\mu\nu}F_{\nu\rho} - F^{\mu\nu}\,^*G_{\nu\rho}+ ^*G^{\mu\nu}F_{\nu\rho} + ^*G^{\mu\nu}\,^*G_{\nu\rho}\,.
\end{eqnarray}
This Lagrangian is manifestly Lorentz covariant, and 
${L}_\mathtt{PST}\rightarrow\mathcal{L}_\mathtt{PST}$ under transformations of Eq.~(\ref{dual:FG}) - invariance under duality transformations.  
The Euler-Lagrangian equations associated to $\mathcal{L}_\mathtt{PST}$ are\cite{Manta:2020wqk}:
\begin{equation}
\partial_\mu F^{\mu\nu}=0\,,\mathrm{and:}\;\partial_\mu G^{\mu\nu}=0\,,    
\end{equation}
where $F^{\mu\nu}$ and $G^{\mu\nu}$ are not independent, but related by the conditions contained in the Lagrangian:
\begin{equation}
\label{dual_cond_PST}
G_{\mu\nu}+^*F_{\mu\nu}=0\,,\mathrm{and:}\;F_{\mu\nu}-^*G_{\mu\nu}=0\,,    
\end{equation}
corresponding to the relations derived in  Eq.~(\ref{*F}). 
Note that in discussing  $\mathcal{L}_\mathtt{PST}$ we assume $\mu_0=\epsilon_0=c=1$ and 
restrict ourselves to Minkowski spacetime.
Under duality transformations:
\begin{eqnarray}
\delta\mathcal{L}_\mathtt{PST}&=&\frac{\partial\mathcal{L}_\mathtt{PST}}{\partial\left(\partial_\mu A_\nu\right)}\, \delta\left(\partial_\mu A_\nu\right)
+\frac{\partial\mathcal{L}_\mathtt{PST}}{\partial\left(\partial_\mu C_\nu\right)}\, \delta\left(\partial_\mu C_\nu\right)\nonumber\\
&=&\partial_\mu\left[-\frac{1}{2}F^{\mu\nu}\,\delta A_\nu
-\frac{1}{2}G^{\mu\nu}\,\,\delta C_\nu \right]\,,
\end{eqnarray}
where we have used the two Euler-Lagrangian equations, and the relations Eq.~(\ref{dual_cond_PST}). 
Since the Lagrangian $\mathcal{L}_\mathtt{PST}$ is invariant under duality transformations,
$\delta\mathcal{L}_\mathtt{PST}=0$, therefore we have: 
\begin{equation}
\partial_\mu\left[-\frac{1}{2}F^{\mu\nu}\,\delta A_\nu
-\frac{1}{2}G^{\mu\nu}\,\,\delta C_\nu \right]=0\,.
\end{equation}
The electromagnetic helicity density is obtained considering the time component and the definition of duality transformations, see Eq.~(\ref{dual:pot}):
\begin{eqnarray}
h^0_\mathtt{em}&=&-\frac{1}{2}F^{0\nu}\,\delta A_\nu
-\frac{1}{2}G^{0\nu}\,\,\delta C_\nu\\
&=&-\frac{1}{2}\eta^{0\sigma}F_{\sigma\nu}\,\delta A^\nu
-\frac{1}{2}\eta^{0\sigma}G_{\sigma\nu}\,\,\delta C^\nu\\
&=&\frac{1}{2}\left(-\eta^{0\sigma}F_{\sigma\nu}C^\nu +  \eta^{0\sigma}G_{\sigma\nu} A^{\nu}  \right) \\
&=&\frac{1}{2}\left(-F^{0\nu}C_\nu +  G^{0\nu} A_{\nu }\right)\,.
\end{eqnarray}
It conicides with the expression for the electromagnetic helicity density
defined in Section II, see the integrand of Eq.~(\ref{h:em:flat}).
Since $\partial_0 h^0+\partial_i h^i=0$ we easily verify, integrating over the volume, 
that total electromagnetic helicity is conserved: 
\begin{equation}
\frac{\partial}{ \partial t} \int_{\Sigma^3} h^0_\mathtt{em} d^3\mathbf{r}=0\,,
\end{equation}

$h^0_\mathtt{em}$ is the Noether's charge density associated to $\mathcal{L}_\mathtt{PST}$ under duality transformations.
It can be easily generalized to curved spacetimes replacing 
ordinary derivatives $\partial_\mu$ 
with covariant derivatives $\nabla_\mu$
and considering a general metric $g_{\mu\nu}$.
An example of Noether's charge density calculation  associated to 
manifestly duality invariant symmetry for a curved spacetime
can be found in \cite{Agullo:2018nfv}.

\section{Conclusions}

We started this work discussing different definitions of helicity present in literature. 
Following the advice in \cite{Deser:1976iy}, and the more recent refs. \cite{Barnett:2012,Cameron:2014optical,Cameron:2014second,Cameron:2014Noether}, we introduced two independent potential vectors $A^{\mu}$ and $C^{\mu}$, respectively for the magnetic induction $\mathbf{B}$ and the electric induction $\mathbf{D}$, in order to have a Lagrangian function manifestly duality invariant.

The relation with the Stokes parameters and their time evolution were derived starting from the Maxwell equations.
We considered two kinds of violation of helicity conservation: magnetic helicity and electromagnetic helicity. The former is associated with rotation of linear polarization,
the latter with circular polarization.
Electromagnetic helicity has to be considered when we focus on photon chiral anomaly.
In Section IV, we described several invariant Lagrangians of the electromagnetic field. 
These Lagrangians provide the same conserved current associated with the duality transformations.
We highlighted only $\mathcal{L}_\mathtt{PST}$ reproduces Maxwell equations. Its relative density charge is the electromagnetic helicity. 
Therefore, quantum anomalies are associated with non-conservation
of circular polarization \cite{Agullo:2014yqa,Agullo:2016lkj,Agullo:2017pyg,Agullo:2018nfv,Agullo:2018iya,delRio:2020cmv}, 
rather than with rotation of linear polarization \cite{Dolgov:1987xv,Reuter:1987ju,Reuter:1991cb}.
Moreover we notice that even
if $R_{\alpha\beta\mu\nu}\,^\star R^{\alpha\beta\mu\nu}\neq0$ locally, 
but the total space integral $\int_{\Sigma^3} R_{\alpha\beta\mu\nu}\,^\star R^{\alpha\beta\mu\nu} d^3\mathbf{r}$ is zero, no measurable effect can be detected as for the Kerr spacetime.  
In order to produce a non-null effect,  we have to consider spacetimes 
with a non-trivial Pontryagin invariant: 
$\int_{\Sigma^3} R_{\alpha\beta\mu\nu}\,^\star R^{\alpha\beta\mu\nu} d^3\mathbf{r}\neq 0$;
this request excludes mirror symmetric spacetimes \cite{delRio:2020cmv}.
The Chern-Pontryagin class is usually non-zero when it is evaluated over certain gravitational instantons, which happens to be solutions of the Euclidean Einstein’s equations only. One way to consider more physical cases, than gravitational instantons, is to study transient phenomena as highlighted in \cite{delRio:2020cmv}.

A future project could be to investigate the electromagnetic field in a chiral medium and the helicity conservation in it. The main point would be to go beyond, both in flat and curved space, the vacuum relations between the electric field induction and the electric field as well as the magnetic induction field and the magnetic field.

\begin{acknowledgements}
We wish to thank Donato Bini, Alfio Bonanno, Martin Reuter, and Dmitri Sorokin
for discussions and valuable comments on this project.
\end{acknowledgements}



\end{document}